\documentclass[prb,twocolumn,floatfix]{revtex4}
\begin{document}
\def\rhov{{\mbox{\boldmath{$\rho$}}}}
\def\tauv{{\mbox{\boldmath{$\tau$}}}}
\def\Lambdav{{\mbox{\boldmath{$\Lambda$}}}}
\def\sigmav{{\mbox{\boldmath{$\sigma$}}}}
\def\xiv{{\mbox{\boldmath{$\xi$}}}}
\def\chiv{{\mbox{\boldmath{$\chi$}}}}
\def\oh{{\scriptsize 1 \over \scriptsize 2}}
\def\ot{{\scriptsize 1 \over \scriptsize 3}}
\def\of{{\scriptsize 1 \over \scriptsize 4}}
\def\tf{{\scriptsize 3 \over \scriptsize 4}}
\title{Erratum: Landau Analysis of the Symmetry of the Magnetic Structure and
Magnetoelectric Interaction in Multiferroics 
[Phys. Rev. B {\bf 76} 054447 (2007)]}

\author{A. B. Harris}

\maketitle

This erratum corrects an error in the anlysis of Sec. IIIC concerning
the allowed Fourier components of the magnetization distribution of
TbMn$_2$O$_5$ for wavevector
${\bf q}=(\oh,0,q_z)$.  The analysis relies on the form of the
inverse susceptibility matrix which was given in Eq. (95).
In constructing the symmetry properties of this matrix, the transformation
under inversion was incorrectly calculated, as is apparent from
Eq. (22). Thus,  in Table XV the entries for inversion ${\cal I}$
and $m_x m_y {\cal I}$ for sublattices 9-12 all were
missing a factor of $\Lambda \equiv \exp(2 \pi i q_z)$, where
$q_z$ is in rlu's.  The consequences of this omission are the
following.  In the matrix of Eq. (95), all the elements $M^{(xx)}_{ij}$
for $i<9$ and $j>8$ should be multiplied by $\Lambda^{1/2}$ and
all the elements of $M^{(xx)}_{i,j}$ with $i>8$ and $j<9$ should be
multiplied by $\Lambda^{-1/2}= {\Lambda^*}^{1/2}$.  However, as before,
the Roman letters are real and the
Greek ones are complex. Typically this form of the matrix arises as
follows.  Using the glide operations $m_x$ and $m_y$ one can show
that $M^{xx}_{1,9}=M^{xx}_{2,10}=M^{xx}_{3,11}= M^{xx}_{4,12}=\alpha$,
where $\alpha$ is an unspecified complex number.  Then inversion
symmetry indicates that $\alpha=M^{xx}_{1,9} = \Lambda [M^{xx}_{10,2}]^*=
\Lambda\alpha^*$. Therefore $\alpha = \Lambda^{1/2} a$, where $a$ is
a real number.

Following this, all the symmetry adapted coordinates
of Eqs. (101) and (104) which pertain to sublattices 9-12 should be
multiplied by $\Lambda^{-1/2}$.  The argument starting with Eq. (112)
and ending with Eq. (118) has missing factors of $\Lambda$, but 
these cancel when Eq. (118) is reached, so that Eq. (119) remains
true, but the ${\bf O}$'s involve Eqs. (101) and (104), modified
as mentioned above.  Then the result for the allowed
spin Fourier transforms of Table XVI is modified by having all the
entries for sublattices 9-12 multiplied by $\Lambda^{-1/2}$.
Finally, the explicit results for the wavefunctions of sublattices
9-12 in Eq. (123) should be corrected by replacing (only for these
sublattices) $\sigmav_n$ by 
$\sigmav_n \Lambda^{-1/2} \equiv \sigma_n \exp(-i \phi_n - i \pi q_z)$.
In Eq. (123) this is most easily done by replacing $\phi_n$ by
$\phi_n + \pi q_z$ for sublattices 9-12.
The details of the argument below Eq. (126) are
changed but the conclusion is still that inversion fixes the phases in the
wavefunction, although now in a way involving $q_z$, which is reasonable.

The reader might well ask, "what makes you think the results are
now correct, since the calculation was complicated?"  The answer is
that this error was found by a calculation\cite{PRE} which shows how
the eigenfunctions of the two 1 dimensional irreducible representations
(for $q_x \not= 1/2$) connect (as $q_x \rightarrow 1/2$)
to these of the 2 dimensional irreducible
representation for $q_x=1/2$.  The equations which gives the
wavefunctions and order parameter of the latter phase in terms of those
of the former phase are highly overdetermined
and the fact that they had no solution using the incorrect wavefunction
(before this erratum) led me to find the error noted in this erratum.
Using the corrected information of Table XVI not only led to
equations which had a solution, but that solution preserved the
phase relations expected from the symmetry analysis leading to
Table XVI.  This result provides strong evidence
that indeed this erratum is conclusive.  Note
that this erratum does not affect the wavefunctions for the Mn
sublattices 1-8, so the results for YMn$_2$O$_5$ do not need
correction.  (The nonmagnetic Y ions occupy sublattices 9-12.)
Indeed my results give an excellent reproduction of the
spin structure of the commensurate phase ${\bf q}=(1/2,0,1/4$)
of YMn$_2$O$_5$, as determined by neutron diffraction.\cite{KIMURA}


\begin{thebibliography} {99}
\bibitem{ABH}
A. B. Harris, Phys. Rev. {\bf 76}, 054447 (2007).
\bibitem{PRE}
A. B. Harris, M. Kenzelman, A. Aharony, and O. Entin-Wohlman, 
unpublished.
\bibitem{KIMURA}
H. Kimura, S. Kobayashi, Y. Fukuda, T. Osawa, Y. Kamada, Y. Noda, I.
Kagomiya, and K. Kohn, J. Phys. Soc. Jpn. {\bf 76}, 074706 (2007).
\end{thebibliography}
\end{document}